\documentclass[twocolumn,openany,openright,amsmath,amssymb,superscriptaddress,prb]{revtex4-2}

\usepackage{float}
\usepackage{latexsym} 
\usepackage{amsmath,amsthm}
\usepackage{ifpdf}
\usepackage{epstopdf}
\usepackage{soul}
\usepackage{dcolumn}
\usepackage{bm}
\usepackage{braket}
\usepackage{wrapfig}
\usepackage{comment}
\usepackage[abs]{overpic}
\usepackage{graphicx,graphics}
\usepackage{dcolumn}
\usepackage{wasysym}
\usepackage{latexsym,verbatim}
\usepackage{color}
\usepackage[framemethod=default]{mdframed}
\usepackage[breaklinks=true,colorlinks,citecolor=blue,linkcolor=blue,urlcolor=blue]{hyperref}
\usepackage[makeroom]{cancel}
\usepackage{fancybox}
\usepackage{tikz-feynman}

\usepackage[caption=false]{subfig}

\DeclareMathOperator{\sign}{sign}

\newcommand{\qq}{\mathbf{q}}

\newcommand{\R}{\mathbf{R}}
\renewcommand{\r}{\mathbf{r}}
\renewcommand{\k}{\mathbf{k}}
\renewcommand{\a}{\mathbf{a}}
\newcommand{\g}{\mathbf{g}}
\newcommand{\p}{\mathbf{p}}
\newcommand{\q}{\mathbf{q}}

\renewcommand{\selectlanguage}[1]{}

\begin{document}
\def \k{\bm k}
\def \l{\bm l}
\def \i{\bm i}
\def \p{\bm p}
\def \ppi{\bm \pi}
\def \q{\bm q}
\def \r{\bm r}
\def \s{\bm s}
\def \P{\bm P}
\def \R{\bm R}
\def \qq{\bm q}
\def \A{\bm A}
\def \D{\bm D}
\def \beq{\begin{equation}}
\def \eeq{\end{equation}}
\def \beal{\begin{aligned}}
\def \eal{\end{aligned}}
\def \bes{\begin{split}}
\def \ees{\end{split}}
\def \besu{\begin{subequations}}
\def \esu{\end{subequations}}
\def \g{\gamma}
\def \G{\Gamma}
\def \ac{\alpha_c}
\def \barr{\begin{eqnarray}}
\def \earr{\end{eqnarray}}

\title{Constructing a variational ground state of matter fermions coupled to a vison pair in Kitaev's honeycomb model}
\author{Theo Grace}
\email{theo.grace@manchester.ac.uk}
\author{Alessandro Principi}
\email{alessandro.principi@manchester.ac.uk}
\affiliation{Department of Physics and Astronomy, University of Manchester, Manchester M13 9PL, UK}
\begin{abstract}
We develop a new method to construct simple and explicit variational approximations for the ground state of Kitaev’s honeycomb model with a non-trivial $\mathbb{Z}_2$ flux configuration consisting of a pair of visons on neighbouring plaquettes. The method consists of retaining only the largest singular values of the generator of the transformation between the vison-pair and flux-free ground states. We compare physical quantities calculated using the approximate state to those obtained by extrapolating results of exact diagonalisation of finite lattices, finding them to be in very good agreement. We discuss ways to extend the method to more complicated flux configurations. 
\end{abstract}
\maketitle
\section{Introduction}

Kitaev’s honeycomb model \cite{kitaev_anyons_2006} is a much-studied exactly solvable model of a frustrated quantum magnet which realises a quantum spin liquid (QSL). It exhibits the characteristic features of a QSL, with long-range entanglement of the ground state and fractionalised excitations \cite{savary_quantum_2017,knolle_field_2019}. In Kitaev’s model the spins fractionalise into Majorana fermions interacting with an emergent $\mathbb{Z}_2$ gauge field, which gives rise to both gapless fermionic excitations (spinons) and gapped excitations due to $\pi$ fluxes of the $\mathbb{Z}_2$ gauge field (visons).

The appeal of Kitaev’s model is not solely for theoretical study of exotic phenomena. There are a growing number of materials for which Kitaev interactions are thought to be dominant \cite{jackeli_mott_2009,trebst_kitaev_2022}, notably $\alpha\text{-RuCl}_3$ \cite{banerjee_proximate_2016,wang_range_2020}. However, despite evidence of strong Kitaev interactions, these materials are also found to magnetically order at low temperatures \cite{johnson_monoclinic_2015,sears_magnetic_2015,chaloupka_zigzag_2013}. This has motivated extensive theoretical study of the interplay between the Kitaev Hamiltonian and other interactions expected in real materials, the largest of which are Heisenberg and symmetric off diagonal (Gamma) interactions \cite{gohlke_dynamics_2017}. The inclusion of other terms in the Hamiltonian spoils its exact solution and produces a rich phase diagram with an extended Kitaev QSL phase as well as a number of competing conventionally ordered phases. This has been established through multiple techniques including parton mean field theories \cite{knolle_dynamics_2018,gohlke_dynamics_2017}, Monte-Carlo methods \cite{wang_one_2019}, coupled chain analysis \cite{gohlke_extended_2022}, and exact diagonalisation \cite{chaloupka_kitaev-heisenberg_2010}. 

An open area of research is finding signatures that can distinguish the QSL from competing phases in experiments. A means of detecting the emergent gauge field would be a valuable indicator of QSL behavior. Several recent theoretical works with this aim have focused on the dynamics of vison excitations in weakly perturbed Kitaev models \cite{zhang_variational_2021,zhang_theory_2022,chen_nature_2023,joy_dynamics_2022,joy_gauge_2024}. The visons, which are static at the exactly solvable point \cite{kitaev_anyons_2006}, become mobile when other interactions are present. The motion of visons sets bounds on the stability of the spin liquid phase which are dependent on the sign of the Kitaev coupling \cite{joy_dynamics_2022}. In magnetic fields visons may form topologically non-trivial bands for antiferromagnetic Kitaev coupling and contribute to a quantized thermal Hall effect \cite{joy_dynamics_2022,chen_nature_2023}. The important role of $\mathbb{Z}_2$ fluxes in phases at higher magnetic field strengths is an active area of research, with recent work suggesting $\mathbb{Z}_2$ flux dynamics produces an emergent glassy phase \cite{Emergent_glassiness} at intermediate field strengths. In weakly perturbed models pairs of visons on neighbouring plaquettes can form bound states with spinons when Heisenberg and Gamma interactions are considered, which can destabilise the spin liquid for sufficiently strong interaction, and has a distinct signature in the spin structure factor \cite{zhang_variational_2021}. The method developed in Ref.~\cite{zhang_variational_2021} relates hopping parameters for vison pairs, generalised to individual visons in \cite{joy_dynamics_2022,chen_nature_2023}, to matrix elements between exact eigenstates of the unperturbed Kitaev model with a fixed gauge. These can then be calculated numerically using exact diagonalisation on finite lattices and extrapolated to infinite ones. 

In order to develop an understanding of vison excitations beyond numerical work it is useful to find simple approximate explicit expressions for Kitaev eigenstates in the presence of vison excitations. The strong interaction between visons and spinons affects many vison properties, such as their excitation energy and mobility in perturbed Kitaev models, which in turn affects the stability of the spin liquid phase itself.  In this work we calculate an approximate explicit expression for the lowest energy eigenstate of the Kitaev model with a non-trivial $\mathbb{Z}_2$ gauge field consisting of a pair of visons on neighbouring plaquettes. We show that a simple ansatz may be used to approximate the vison pair state which we then optimise by minimising the matter sector ground state energy.
We then compare results for hopping parameters of vison pairs obtained with our approximate method to those of Ref.~\cite{zhang_variational_2021} based on exact diagonalisation. Our results are summarised in Table~\ref{tab:table1}. We note that the largest relative error is of the order of just $\approx 6\%$, which shows the good performance of the method we have developed. We stress that, compared to exact diagonalisation for which the largest systems studied are of the order $100\times100$ unit cells, our method allows direct study of much larger lattices exceeding $1000\times1000$ unit cells.

\begin{table}
\caption{\label{tab:table1}%
Comparison of vison pair hopping amplitudes calculated using exact diagonalisation on finite lattices extrapolated to infinite size and the variational approximation.
}
\begin{ruledtabular}
\begin{tabular}{cccc}
\textrm{Interaction}&
\textrm{Kitaev sign}&
\textrm{Exact}&
\textrm{Approximate}\\
\colrule
Heisenberg & FM & 0.0938$J$\footnote{Taken from \cite{zhang_variational_2021}} & 0.0991$J$\\
& AFM & 1.4702$J$\footnotemark[1] & 1.4964$J$ \\
Gamma & FM & -1.4391$\Gamma$\footnotemark[1]  & -1.4632$\Gamma$\\
& AFM & -0.1733$\Gamma$\footnotemark[1] & -0.1809$\Gamma$ \\
Zeeman & FM  & 1.35i $h$\footnote{Taken from \cite{zhang_theory_2022}.} & 1.387i $h$ \\
& AFM & 0.0494i$h$\footnotemark[2] & 0.0495i$h$
\end{tabular}
\end{ruledtabular}
\end{table}

The structure of the paper is as follows. In section \ref{sec:Review_Kitaev} we review Kitaev’s honeycomb model and its exact solution. In section~\ref{sec:vison_pair_state} we provide a justification for the variational ansatz and for the application of the variational method to the determination of approximate eigenstates. In section~\ref{sec:Comparison_to_exact_results} we compare the results from our approximation with those obtained via exact diagonalisation. Finally, in section \ref{sec:discussion} we discuss our results and draw conclusions, highlighting possible further applications of our method.

\section{Review of Kitaev's honeycomb model}\label{sec:Review_Kitaev}
\begin{figure}
    \centering
    \includegraphics[width=8.5cm]{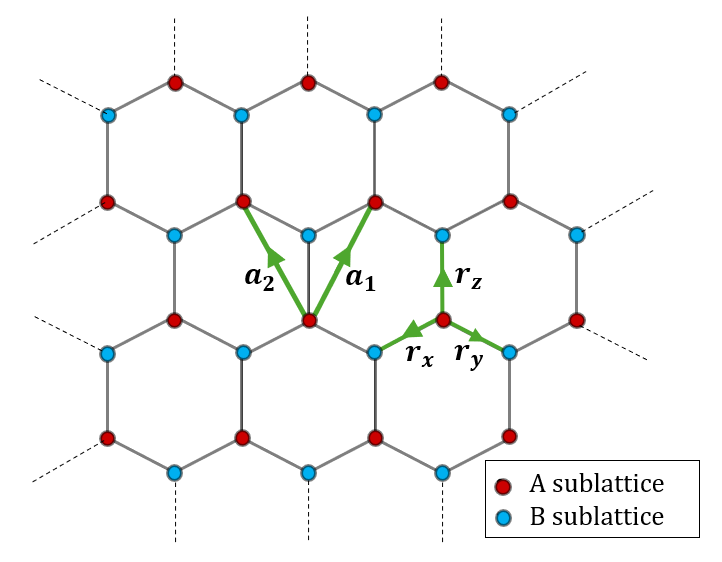}
    \caption{The geometry of Kitaev's honeycomb model. $\a_1 = \frac{1}{2}(1,\sqrt{3})$ and $\a_2=\frac{1}{2}(-1,\sqrt{3})$ are the primitive lattice vectors, while $\r_\alpha$ ($\alpha=x,y,z$) denotes the vectors connecting a site of sublattice A to its three B-type nearest neighbours. Dimensionless units are used such that $|\a_1|=|\a_2|=1$.}
    \label{fig:Model_Geometry}
\end{figure}

Kitaev's honeycomb model describes spins $S=1/2$ on a honeycomb lattice with a bond direction dependent Ising interaction between nearest neighbour spins \cite{kitaev_anyons_2006}. The Hamiltonian is given by: 
\begin{align}
    \mathcal{H} = -K\sum_{\r,\alpha} \sigma^\alpha_{\r}\sigma_{\r+\r_\alpha}^\alpha,
\end{align}
where $\sigma^\alpha$ ($\alpha= x,y,z$) are Pauli matrices, $\r$ labels a site on the A sublattice,  while $\r_\alpha$ are the vectors connecting such site to its three nearest neighbours (which belongs to the B sublattice).
See Figure \ref{fig:Model_Geometry} for the definition of $\r_\alpha$.

This Hamiltonian has an extensive number of conserved quantities which are crucial for its exact solution~\cite{kitaev_anyons_2006}. For each plaquette one can define a (conserved) ``flux'' operator as the product of Pauli matrices associated to exterior bonds~\cite{kitaev_anyons_2006} [see also Eq.~(\ref{eq:flux_definition}) below].
The model can be exactly solved using Kitaev's representation of each spins in terms of four Majorana fermions, named $b_{\r}^{x}$, $b_{\r}^{y}$, $b_{\r}^{z}$ and $c_{\r}$, such that $\sigma_{\r}^{\alpha} = ib_{\r}^{\alpha} c_{\r}$. We will take the convention that the Majorana operators satisfy $\{c_{\r},c_{\r'}\} = 2\delta_{\r,\r'}$, $\{b_{\r}^{\alpha},b_{\r'}^{\beta}\} = 2\delta_{\r,\r'}\delta_{\alpha,\beta}$, $\{c_{\r},b_{\r}^{\alpha}\} = 0$. Complex fermions formed from combinations of Majoranas denoted as $c_{\r}$ will be referred to as 'matter' fermions. The Majorana representation has a larger Hilbert space than that of the spin states: only 
states which satisfy $D_{\r}\ket{\psi} = \ket{\psi}$ with $D_{\r} = b_{\r}^{x}b^{y}_{\r} b^{z}_{\r} c_{\r}$ represent physical spin states.
The states in the enlarged Hilbert space can be projected onto the physical subspace by using the projection operator: 
\begin{equation} \label{eq:proj_op}
    \hat{P} =\prod_{\r} \frac{1+D_{\r}}{2},
\end{equation}
The Hamiltonian in the Majorana representation is given by: 
\begin{align}
    \mathcal{H}^{\hat{u}} = iK\sum_{\r\in A,\alpha} c_{\r} \hat{u}_{\r}^{\alpha} c_{\r+\r_\alpha},
\end{align}
where $\hat{u}^{\alpha}_{\r} = ib_{\r}^{\alpha} b_{\r+\r_\alpha}^{\alpha}$ is a (conserved) link operator with eigenvalues $\pm1$. 
The link operators satisfy $[\hat{u}_{\r}^{\alpha},\hat{u}_{\r'}^{\beta}] = 0$ and $[\hat{u}_{\r}^{\alpha},\mathcal{H}^{\hat{u}}]=0 $, so the eigenstates may be labelled by their link operator eigenvalues. 
This allows eigenstates in the extended Hilbert space to be written as a tensor product of the state of the link variables denoted as $\ket{u}$ and the state of the matter fermions. 
The Hamiltonian is invariant under local $\mathbb{Z}_2$ transformations in which $c_{\r} \rightarrow (-1)^{n_{r}}c_{\r}$, $\hat{u}_{\r}^{\alpha} \rightarrow (-1)^{n_{\r}}\hat{u}_{\r}^{\alpha}(-1)^{n_{\r+\r_\alpha}}$ with site dependent $n_{\r} \in \{0,1\}$. A Wilson loop given by a  product of link operators $\hat{u}_{\r}^{\alpha}$ around a closed loop is invariant under a $\mathbb{Z}_2$ gauge transformation, and corresponds to the flux of the emergent $\mathbb{Z}_2$ gauge field through the loop. The smallest Wilson loop is around a single hexagonal plaquette: 
\begin{equation} \label{eq:flux_definition}
    W_p = \prod_{u^{\alpha} \in \partial \varhexagon_{p}} \hat{u}_{\r}^{\alpha}.
\end{equation}
The loop operators commute with the Hamiltonian and satisfy $[W_p,D_{\r}]=0$ so 
physical states may be labeled by their $\mathbb{Z}_2$ flux configuration. 

Acting on an eigenstate of the link operators $\ket{u}$ gives a  Hamiltonian for the matter fermions which is quadratic in matter fermion operators. 
\begin{align} \label{eq:Matter_sector_Hamiltonian}
    \mathcal{H}^{u} &= iK\sum_{\r,\r' \in A} c_{\r}^{A}M_{\r,\r'}^{u}c_{\r'}^{B}.
\end{align}
Here $M^{u}_{\r,\r'} = u^{z}_{\r} \delta_{\r,\r'}+u^{x}_{\r} \delta_{\r-\a_1,\r'}+{u}^{y}_{\r} \delta_{\r-\a_2,\r'}$ is a link variable dependent matrix and the Majorana operators have been given explicit sublattice labels $c^{A}_{\r} = c_{\r}$ and $c^{B}_{\r} = c_{\r+\r_z}$.
This may be diagonalised exactly by taking the singular value decomposition $M^{u}= U^{u}E^{u}V^{uT}$, where $U^{u}$ and $V^{u}$ are real orthogonal matrices and $E^{u}$ is a diagonal matrix of singular values. The superscript $u$ is used to highlight the dependence on the configuration ({\it i.e.} the eigenstates) of the link operators. Note that although the singular values $E_{k}^{u}$ depend on the link configuration, they are independent of $\mathbb{Z}_2$ gauge transformations, so they only depend on the $\mathbb{Z}_2$ flux configuration. Using the singular value decomposition, one can rewrite the Hamiltonian: 
\begin{align} \label{eq:Ham_SVD}
    \mathcal{H}^{u} 
    &= iK\sum_{k} E^{u}_{k} C^{u,A}_{k}C^{u,B}_{k} \nonumber\\
    &= |K|\sum_{k}E_{k}^{u}(2\eta_{k}^{u\dagger}\eta_{k}^{u}-1),
\end{align}
where $C_{k}^{u,A} = U^{u,T}_{k,\r}c_{\r}^{A}$ and $C_{k}^{u,B} = V_{k,\r}^{u,T}c_{\r}^{B}$ are Majorana fermions which may be combined into a complex fermion which diagonalises the matter sector Hamiltonian for a particular link variable configuration. 
\begin{equation}\label{eq:Kitaev_matter_eigenstate}
    \eta_{k}^{u} = (C_{k}^{u,A}+ i K/|K|C_{k}^{u,B})/2
\end{equation}
Note that we have included a factor which depends on the sign of the Kitaev coupling in the definition of $\eta_{k}^u$ to ensure $\eta_{k}^u$ is an annihilation operator for both ferromagnetic and antiferromagnetic Kitaev interaction. The matter fermion ground state corresponding to a particular link variable configuration is then defined by: 
\begin{align*}
    \eta_{k}^u\ket{0^u} = 0 \quad \forall k. 
\end{align*}
Lieb's Theorem \cite{lieb_flux_1994} guarantees that the ground state of the model is in the fluxless sector in which the Wilson loop operator has eigenvalues $W_{p} = +1$ on all plaquettes. The simplest link configuration  in this sector is the trivial configuration, labelled by $u_{0}$,  in which all link variables are $u_{\r}^{\alpha}=+1$. For this configuration the corresponding matter sector Hamiltonian is a translationally invariant nearest neighbour hopping model, so the diagonalising matter fermion operators may be found explicitly using a Fourier transformation \cite{kitaev_anyons_2006}. 
\begin{align}\label{eq:Fluxless_eigenstates}
    \eta_{\k}^{u_0} &= \frac{1}{\sqrt{2N}}\sum_{\r}\left(e^{i\k\cdot\r}c_{\r}^{A} + i\frac{K}{|K|}e^{i(\k\cdot\r+\varphi_{\k})}c_{\r}^{B}\right)
\end{align}
Where $\varphi_{\k}$ is defined by $e^{i\varphi_{\k}} = \Delta_{\k}/|\Delta_{\k}|$ with $\Delta_{\k} = 1 + e^{i\k\cdot\a_1} + e^{i\k\cdot\a_2}$. The corresponding energies are $E_{\k}^{u_{0}} = 2|K||\Delta_{\k}|$. Note that the operators in Eq.~(\ref{eq:Fluxless_eigenstates}) are not written as the sum of a Majorana operator on each sublattice in the form of Eq.~(\ref{eq:Kitaev_matter_eigenstate}). The operators $\eta_{\k}^{u_0\dagger}$ and $\eta_{-\k}^{u_0\dagger}$ create degenerate excitations so may be combined into new eigenstate creation operators $\eta_{\k,+}^{u_0}=(\eta_{\k}^{u_{0}}+\eta_{-\k}^{u_{0}})/\sqrt{2}$ and $\eta_{\k,-}= i(\eta_{-\k}^{u_{0}}-\eta_{\k}^{u_{0}})/\sqrt{2}$ which do have the form of Eq.~(\ref{eq:Kitaev_matter_eigenstate}).
The physical ground state is given by: 
\begin{align}
    \ket{\Omega} = \hat{P}(\ket{u_{0}} \otimes \ket{0^{u_{0}}}).
\end{align}
As well as the matter fermion excitations, there are excitations associated with link variable configurations in which there are Wilson loops with $W_{p}=-1$ on some of the plaquettes. These excitations, due to a flux of the emergent $\mathbb{Z}_2$ gauge field, are called visons.

\section{Matter fermion ground state in the presence of a vison pair}\label{sec:vison_pair_state}
A $\mathbb{Z}_2$ flux configuration with a pair of visons on neighbouring plaquettes may be represented by a single link variable $u_{\r}^{\alpha} = -1$ with all others equal to +1. The position of a vison pair can be labelled by the link joining the plaquettes, which can be identified by a site $\r$ on the A sublattice and the link orientation $\alpha$. For a vison pair at the origin separated by a $z$-oriented link, the link state may be written as $b_0^z\ket{u_{0}}$ and the corresponding Hamiltonian for the matter fermions is $\mathcal{H}^{0,z}$, with ground state $\ket{0^{0,z}}$. The Hamiltonian $\mathcal{H}^{0,z}$ can be written as the sum of the trivial link configuration Hamiltonian $H^{u_{0}}$  and a local perturbation due to the flipped link variable $u_{0}^z$:
\begin{align}
    \mathcal{H}^{0,z} &=  \mathcal{H}^{u_{0}} - 2iKc_0^Ac_{0}^B. 
\end{align}
The Hamiltonian with a single flipped link $\mathcal{H}^{0,z}$ can be written in terms of the excitations of the trivial link configuration: 
\begin{align}
    \mathcal{H}^{0,z} &= \mathcal{H}^{u_{0}}+ 2|K|U^{u_{0}}_{0,k}V^{u_{0}}_{0,k'}(\eta_{k}^{u_{0}}+\eta_{k}^{u_{0}\dagger})(\eta_{k'}^{u_{0}\dagger}-\eta_{k'}^{u_{0}})
\end{align}
where a sum over indices $k$ and $k'$ is understood.
Written in this form the Hamiltonian is independent of the sign of the Kitaev coupling. This means that the ground state $\ket{0^{0,z}}$ will have the same form for both signs of the Kitaev interaction,
 when expressed in terms of the creation and annihilation operators $\eta^{u_{0}}$, $\eta^{u_{0}\dagger}$ and the ground state of the trivial flux configuration $\ket{0^{u_{0}}}$ [see Eq.~(\ref{eq:GS_relation}) below]. It is important to emphasise that the ground state wavefunction is not the same for both signs of the Kitaev interaction, as the dependence on the sign of K has been included in the definition of $\eta^{u_{0}}$.
The difference between eigenstates for different signs of the Kitaev interaction becomes important when non-Kitaev interactions are included in the Hamiltonian. 

Using the singular value decomposition of $M^{u}$ in Eq.~(\ref{eq:Matter_sector_Hamiltonian}) the Majorana operators $C^{u,A}$ and $C^{u,B}$ on the A and B sublattice which diagonalise each Hamiltonian may be related by real orthogonal transformations: 
\begin{equation*}
    C^{(0,z),A} = UC^{u_{0},A}, \quad \quad C^{(0,z),B} = VC^{u_{0},B},
\end{equation*}
where $U = U^{(0,z),T}U^{u_0}$ and $V = V^{(0,z),T}V^{u_0}$.
The relation between the complex fermions $\eta^{(0,z)}$ and $\eta^{u_{0}}$, defined in Eq.~(\ref{eq:Kitaev_matter_eigenstate}),
is given by a real transformation:
\begin{align}
    \begin{pmatrix}
        \eta^{(0,z)} \\ \eta^{(0,z) \dagger}
    \end{pmatrix} = 
    \begin{pmatrix}
        X & Y \\ Y & X 
    \end{pmatrix}
    \begin{pmatrix}
        \eta^{u_{0}} \\ \eta^{u_{0} \dagger}
    \end{pmatrix},
\end{align}
where $X = (U+V)/2$ and $Y=(U-V)/2$. The corresponding matter sector ground states 
are related by \cite{ripka_quantum_1986}: 
\begin{align} \label{eq:GS_relation}
    \ket{0^{0,z}} &= \braket{{0^{u_{0}}}|{0^{0,z}}}\exp\left(-\frac{1}{2}\sum_{k,k'}\eta^{u_{0}\dagger}_{k} Z_{k,k'}\eta^{u_{0}\dagger}_{k'}\right)\ket{0^{u_{0}}},
\end{align}
where $Z = X^{-1}Y$ is a real antisymmetric matrix. As $Z$ is real antisymmetric it may be decomposed by a real orthogonal matrix $Q$ into the form: 
\begin{align}
    Z = Q
    \begin{pmatrix}
        &-t_1&&&&  \\
        t_1&&&&&\\
        &&&-t_2&& \\
        &&t_2&&& \\
        &&&&\ddots&
    \end{pmatrix}Q^{T},
\end{align}
where the non-zero elements $t_j = \tan(\theta_{j}/2)$ are real numbers, and $|\tan(\theta_{j}/2)|$ correspond to the singular values of $Z$. The orthogonal matrix $Q$ can be used to define a canonical transformation to a new set of creation operators $\Bar{\eta}_j^\dagger = Q_{jk}^T\eta_{k}^\dagger$. 
The transformed creation operators corresponding to non-zero $\tan(\theta_{j}/2)$ are paired, with the two operators in the pair labelled $\Bar{\eta}^{u_{0}\dagger}_{j,a}$, $ \Bar{\eta}^{u_{0}\dagger}_{j,b}$. They are not unique, as the following transformation leaves the relation between ground states unchanged:
\begin{align}\label{eq:U(1)_redundancy}
    \begin{pmatrix}
        \Bar{\eta}_a^{\dagger} \\ \Bar{\eta}_b^\dagger 
    \end{pmatrix}
    \rightarrow \exp\left(i\phi\sigma^y\right)
    \begin{pmatrix}
        \Bar{\eta}_a^{\dagger} \\ \Bar{\eta}_b^\dagger 
    \end{pmatrix},
    \quad \phi \in \mathbb{R}.
\end{align}
The operator relating ground states may then be written in terms of the transformed operators as
\begin{align}
    \exp\left(-\frac{1}{2}\sum_{k,k'}\eta^{u_{0}\dagger}_{k} Z_{k,k'}\eta^{u_{0}\dagger}_{k'}\right)
    &= \exp\left(\sum_j t_j \Bar{\eta}^{u_{0}\dagger}_{j,a} \Bar{\eta}^{u_{0}\dagger}_{j,b} \right) \nonumber\\
    = \prod_j (1 + \tan(\theta_{j}/2)&\Bar{\eta}^{u_{0}\dagger}_{j,a} \Bar{\eta}^{u_{0}\dagger}_{j,b}).
    \label{eq:Product_over_pairs}
\end{align}
Here, sums and products are over the pairs of non-zero singular values of $Z$. 

A single flipped bond variable is a local perturbation of the flux free Hamiltonian $\mathcal{H}^{u_{0}}$ and the change to the ground state as described by $Z$ is localised around the vison pair \cite{zhang_variational_2021}. This means that there are very few non-zero values of $\tan(\theta_{j}/2)$, with the corresponding fermion operators localised around the vison pair. The change to the ground state may be approximated as being due solely to the largest term such that the operator relating the ground states of the two sectors is: 
\begin{align}\label{eq:EXP_approximation}
   \ket{0^{0,z}} &= \braket{{0^{u_{0}}}|{0^{0,z}}}\exp{\left(-\frac{1}{2}\sum_{k,k'}\eta^{u_{0}\dagger}_{k} Z_{k,k'}\eta^{u_{0}\dagger}_{k'}\right)}\ket{0^{u_{0}}} \\
   &\approx (\cos(\theta/2) + \sin(\theta/2)\Bar{\eta}^{u_{0}\dagger}_{a} \Bar{\eta}^{u_{0}\dagger}_{b})\ket{0^{u_{0}}}.
\end{align}
The matrix $Z$ may be determined exactly numerically for finite size systems using exact diagonalisation of the matter sector Hamiltonians for the two different flux configurations, as has been carried out in previous works on vison dynamics in perturbed Kitaev models \cite{zhang_variational_2021,zhang_theory_2022,chen_nature_2023,joy_dynamics_2022}. However by approximating $Z$ by only its largest singular values the relation between ground states is simple enough to find an approximate explicit expression. This allows calculations on much larger lattices than can be directly studied using exact diagonalisation, and is a simple starting point for further study of vison pair excitations in perturbed Kitaev models. The approximation in Eq.~(\ref{eq:EXP_approximation}) will be the ansatz for a variational calculation of the matter fermion ground state in the presence of a neighbouring pair of visons. 

\subsection{Variational approximation of the vison pair state}
The ansatz for the ground state of the matter sector Hamiltonian for a single flipped bond is: 
\begin{align}\label{eq:pair_variational_ansatz}
    \ket{\Bar{0}^{0,z}}  = \left(\cos(\theta/2) + \sin(\theta/2)\Bar{\eta}_{a}^\dagger\Bar{\eta}_{b}^\dagger\right)\ket{0^{u_{0}}}
\end{align}
The operator $\Bar{\eta}_a^\dagger$ can be written in the basis of the eigenstate operators of $\mathcal{H}^{u_{0}}$: 
\begin{align*}
    \bar{\eta}_a^\dagger  &= \sum_{\k \in BZ/2} (a_{\k,+}\eta_{\k,+}^{u_{0},\dagger} + a_{\k,-}\eta_{\k,-}^{u_{0},\dagger})
    = \sum_{\k \in BZ} a_{\k} \eta_{\k}^{u_{0},\dagger} \\
    \bar{\eta}_b^\dagger  &= \sum_{\k \in BZ/2} (b_{\k,+}\eta_{\k,+}^{u_{0},\dagger} + b_{\k,-}\eta_{\k,-}^{u_{0},\dagger})
    = \sum_{\k \in BZ} b_{\k} \eta_{\k}^{u_{0},\dagger}
\end{align*}
where $a_{\k,+}$, $a_{\k,-}$ are real and $a_{\k} = (a_{\k+}+ia_{\k,-})/\sqrt{2}$ are complex and satisfy $a_{-\k} = a_{\k}^{*}$. The expansion coefficients $b_{\k}$ satisfy analogous conditions. The  eigenstate operators $\eta_{\k}^{u_{0}}$ from Eq.~(\ref{eq:Fluxless_eigenstates}) and complex parameters $a_{\k}$, $b_{\k}$ are used in order to make the following calculation more compact. 
The parameters $a_{\k}$, $b_{\k}$, and $\theta$ are to be determined by minimisation of the ground state energy expectation value. Note that the parameters $a_{\k}$, $b_{\k}$ are constrained by their normalisation $\sum_{\k} |a_{\k}|^2 = \sum_{\k} |b_{k}|^2 = 1$ and orthogonality $\sum_{\k} a_{\k} b_{\k}^{*} = 0 $, with the constraints enforced by Lagrange multipliers $\Lambda_{a}$, $\Lambda_{b}$, and $\Lambda_{ab}$ respectively.

We performed the minimisation of the expectation value for the ground state energy $\Bar{E}^{0,z} = \bra{\Bar{0}^{0,z}}\mathcal{H}^{0,z}\ket{\Bar{0}^{0,z}}$ with respect to each of the variational parameters and solve the resulting equations in a case where $\Lambda_{ab}=0$. This is always possible for a suitable value of $\phi$ in Eq. (\ref{eq:U(1)_redundancy}), and this is a convenient choice as the resulting equations for $a_{\k}$ and $b_{\k}$ are decoupled.
The details of the solution are given in appendix~\ref{app:Minimisation}. For the most general case of a vison pair at site $\r$ with orientation $\alpha$ the optimised parameters are: 
\begin{equation}\label{eq:a_k_optimised}
    a_{\k}^{(\r,\alpha)} = \frac{A_a}{\sqrt{N}}\frac{1+e^{i\varphi_{\k}+i\k\cdot(\r_\alpha-\r_z)}}{\Lambda_{a}+|\Delta_{\k}|}e^{i\k\cdot\r},
\end{equation}
\begin{equation}\label{eq:b_k_optimised}
    b_{\k}^{(\r,\alpha)} = \frac{A_b}{\sqrt{N}}\frac{e^{i\varphi_{\k}+i\k\cdot(\r_\alpha-\r_z)}-1}{\Lambda_{b}+|\Delta_{\k}|}e^{i\k\cdot\r}.
\end{equation}
with
\begin{align*}
    \Lambda_{a} = 1.5384, \\
    \Lambda_{b} = 0.2995, \\
    \theta = 0.9565.
\end{align*}
The energy of the vison pair above the ground state in this approximation is given by $\Bar{\Delta} = \Bar{E}^{0,z}-\bra{0^{u_{0}}}\mathcal{H}^{u_0}\ket{0^{u_{0}}} \approx 0.272|K|$, which is within $3\%$ of the exact result calculated analytically $\Delta = 0.263|K|$~\cite{panigrahi_analytic_2023}.

\section{Comparison to exact results}\label{sec:Comparison_to_exact_results}
We evaluate the accuracy of the optimized approximate ground state by calculating physical quantities and comparing them to the results of exact diagonalisation performed on finite lattices~\cite{zhang_theory_2022,zhang_variational_2021}. While the ground state is calculated for a particular link configuration, and it is therefore not gauge invariant, it can still be used to calculate physical quantities by acting on it with the projection operator in Eq.~(\ref{eq:proj_op}). Depending on the boundary conditions \cite{zschocke_physical_2015}, the projection operator will annihilate states with either even or odd fermion parity. In cases in which a lone vison pair is annihilated, the projector may be modified to annihilate states with the opposite parity, as detailed in \cite{zhang_variational_2021}.

\subsection{Link energy around vison pairs}
\begin{figure}[t!]
    \begin{center}
    \begin{tabular}{c}
    \begin{overpic}[width=0.95\columnwidth]{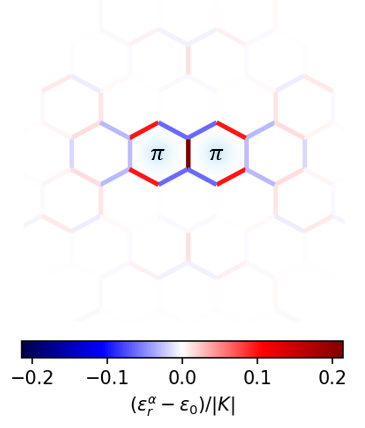}
\put(2,0){{\large (a)}}
\end{overpic}
    \\
    \begin{overpic}[width=0.95\columnwidth]{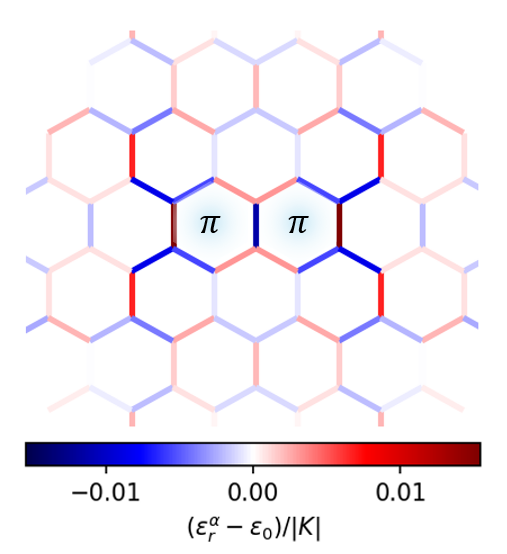}
\put(2,0){{\large (b)}}
\end{overpic}
    \end{tabular}
    \end{center}
    \caption{\label{fig:link_energy} 
    Panel (a): The change in link energy around the vison pair calculated using the variational Ansatz in Eq.~(\ref{eq:pair_variational_ansatz}).
    Panel (b): The difference between the link energies calculated using the variational Ansatz~(\ref{eq:pair_variational_ansatz}) and exact diagonalisation on a $58\times58$ lattice with periodic boundaries. Note the change in scale compared to Fig.~(\ref{fig:link_energy}) by about one order of magnitude.
    In both panels, visons are located on the shaded plaquettes marked with ``$\pi$''.}
\end{figure}

We start by calculating the change in link energy, defined as $\varepsilon_{\r}^{\alpha}= K\langle \sigma_{\r}^{\alpha} \sigma_{\r+\r_{\alpha}}^{\alpha}\rangle = K\langle \hat{u}_{\r}^{\alpha}i c_{\r} c_{\r+\r_\alpha}\rangle$, between the ground states with and without the vison pair. This quantity provides a direct test for the performance of our optimized approximation.
Fig.~\ref{fig:link_energy}(a) shows the change in link energy calculated using the optimised approximation. The majority of the change in link energy is localised on the $x$ and $y$ links neighbouring the visons, and on the flipped $z$ link itself. The difference between the link energy calculated using exact diagonalisation for a finite lattice of $58\times 58$ unit cells and the optimised approximation is shown in Fig.~\ref{fig:link_energy}(b). Note the difference in scale between Fig.~\ref{fig:link_energy}(a) and~(b). The largest differences are on the $z$ bonds neighbouring the plaquette. In the central region where the majority of the energy is localised the relative error is approximately $5\%$. This allows us to conclude that the approximation is robust and we expect it to be able to produce reliable results. Furthermore, it can be consistently improved by including further terms coming from the second, third, etc. largest singular values of $Z$ in the variational Ansatz~(\ref{eq:pair_variational_ansatz}).  

\subsection{Hopping of vison pairs}

 Zhang et.al \cite{zhang_variational_2021} showed that the hopping parameters for vison pairs can be related to matrix elements of matter sector ground states for fixed link configurations. The hopping parameter of vison pairs in the presence of a perturbing Heisenberg interaction $\mathcal{H}^{J} = J \sum_{\r,\alpha} \sigma_{\r} \cdot \sigma_{\r+\r_\alpha}$ is \cite{zhang_variational_2021}: 
\begin{align}\label{eq:Heisenberg_hopping}
    \mathcal{T}^{J} &= \bra{\r+\a_1,z}\mathcal{H}^{J}\ket{\r,z} \nonumber\\
    &= J(\braket{{0^{\a_1,z}}|{0^{0,z}}}+\bra{0^{\a_1,z}}ic_{\a_1}^{A}c_{0}^{B}\ket{0^{0,z}})
\end{align}
This was calculated using exact diagonalisation on finite size lattices and extrapolating to the limit of infinite-size lattices~\cite{zhang_variational_2021}. The dependence on the sign of the Kitaev coupling of the second term is clear once matrix elements are expressed in terms of creation and annihilation operators of matter fermion excitations for a given link configuration: 
\begin{align*}
    &\bra{0^{\a_1,z}}ic_{\a_1}^Ac_0^B\ket{0^{0,z}}\\
    &= \frac{K}{|K|}U^{u_0}_{\a_1,k}V^{u_0}_{0,k'} \bra{0^{\a_1,z}}(\eta_k^{u_0}+\eta_{k}^{u_0\dagger})(\eta_{k'}^{u_0}-\eta_{k'}^{u_0\dagger})\ket{0^{0,z}}
\end{align*}
Once the matrix elements are expressed using matter fermion excitation operators they are independent of the sign of K, so the second term in Eq.~(\ref{eq:Heisenberg_hopping}) differs in sign between AFM and FM Kitaev models, as concluded in \cite{zhang_variational_2021}. These matrix elements may also be calculated using the approximate ground states with the optimised parameters. The details of the calculation are given in the Appendix~\ref{App:Hopping_parameters}. For antiferromagnetic (AFM) Kitaev coupling our approximation gives $\mathcal{T}^J = 1.496$, compared to $\mathcal{T}^J \rightarrow 1.470J$ from exact diagonalisation~\cite{zhang_variational_2021}. For ferromagnetic coupling, as noted in \cite{zhang_variational_2021}, there is destructive interference between the two terms. Our approximation gives $\mathcal{T}^J = 0.0991J$ whereas the extrapolation of exact diagonalisation results gives $\mathcal{T}^J \rightarrow 0.0938J$. The relative error is approximately $6\%$, which denotes a good performance of the approximate method, particularly as this quantity is sensitive to small changes in the ground state due to the destructive interference between the two terms. 

In appendix we have also calculate hopping due to the off-diagonal symmetric exchange $\mathcal{H}^{\Gamma} = \Gamma\sum_{\r} \sum_{\alpha\neq\beta\neq\gamma}\sigma_{\r}^{\beta}\sigma_{\r+\r_{\alpha}}^{\gamma}$, and the Zeeman coupling $\mathcal{H}^{h} = h\sum_{\r,\alpha}\sigma_{\r}^{\alpha}$. Our results, compared to the ones extrapolated from exact diagonalisation of finite lattices~\cite{zhang_variational_2021,zhang_theory_2022}, are reported in Table~\ref{tab:table1} in units of the respective coupling constants.

\section{Discussion}\label{sec:discussion}

In this work we have shown that the change to the matter fermion ground state due to the presence of a vison pair in a generalised Kitaev model may be approximated by a simple expression. Using the variational method we developed here, we obtained an analytical expression for the vison-pair state in terms of the 
eigenstates of the vison-free Hamiltonian.
 
The approximate state was used to calculate the link energy and vison hopping amplitudes
that
we then compared to results of exact diagonalisation. We have shown that the approximate state is successful in capturing the main effect of the vison pair-spinon interaction to within $\approx 6\%$. A more precise determination of the matrix elements requires including more variational parameters in the ansatz corresponding to the smaller singular values, or resorting to exact diagonalisation.

The power of the method we have developed hinges on the possibility to explicitly construct a simple expression for the vison pair state, Eq.~(\ref{eq:pair_variational_ansatz}). This is achieved by retaining only the contribution of the largest pair of singular values of $Z$ to Eq.~(\ref{eq:Product_over_pairs}). 
More precise results may be obtained by including further terms coming from the second, third, etc. largest pairs of singular values of $Z$ in the variational ansatz. However, this complicates the minimisation and it is beyond the scope of the present work. 

The variational method we have developed could be used 
to go beyond numerical calculations in studies of, e.g., magnon-like bound states formed between vison pairs and spinons or to describe the ground state of the extended Kitaev model~\cite{kitaev_anyons_2006}. The latter includes time-reversal symmetry breaking three-spin interactions that emerge in weak magnetic fields.  
It can also in principle be
extended to ground states of more complicated vison configurations. 
For example to ``open pairs'', in which visons  
are separated by longer chains of flipped link variables. 
The study of these more complicated cases is however beyond the scope of the present work.

If $Z$ presents too many singular values that need to be retained, a direct application of the method could become impractical due to the number of variational parameters needed to accurately describe the state. 
In these cases, a possible solution could be breaking too long chains in shorter ones, {\it i.e.} by
considering changes to the ground state induced by flipping a smaller number of links. These could likely be written in terms of few variational parameters onto which minimisation can be performed. Repeating the process, it could then be possible to build up and approximate the ground state of longer chains of flipped bonds. 
Further work, beyond the scope of the present paper, is needed to generalise our method to these cases.

\acknowledgements 

We acknowledge support from the European Commission under the EU Horizon 2020 MSCA-RISE-2019 programme (project 873028 HYDROTRONICS), from the Leverhulme Trust under the grant agreement RPG-2023-253 and from the Engineering and Physical Sciences Research Council grant number 2856485.

\appendix

\section{Minimisation of  the ground state energy}\label{app:Minimisation}

The optimised parameters for the variational ansatz given in Eq.~(\ref{eq:pair_variational_ansatz}) are found using the standard variational method, {\it i.e.} by writing down the ground state energy expectation in terms of the variational parameters and minimising with respect to them. The expectation value for the ground state energy $\Bar{E}^{0,z} = \bra{\Bar{0}^{0,z}}\mathcal{H}^{0,z}\ket{\Bar{0}^{0,z}}$ in terms of the variational parameters is- given by Eq.~(\ref{eq:GS_energy_expectation}): 
\begin{equation}\label{eq:GS_energy_expectation}
\begin{split}
    \Bar{E}^{0,z}
    &= \bra{0^{u_{0}}} H^{0 z }\ket{0^{u_{0}}} \\
    &+2\sin^2(\theta/2)\sum_{\k} |\Delta_{\k}| (|b_{\k}|^2+|a_{\k}|^2) \\
    &+ 2\sin(\theta)((Ub)_0 (Va)_{0}-(Ua)_{0} (Vb)_0)\\
    &-4\sin^2(\theta/2)((Ua)_{0} (Va)_{0}+(Ub)_0 (Vb)_{0}) \\
    &+ 2\sin^2(\theta/2)\Lambda_{a}(\sum_{\k}|a_{\k}|^2-1) \\
    &+ 2\sin^2(\theta/2)\Lambda_{b}(\sum_{\k}|b_{\k}|^2-1) \\
    &+ \Lambda_{ab}(\sum_{\k} a_{\k} b_{\k}^{*}),
\end{split}
\end{equation}
where
\begin{align*}
    (Ua)_{0} &= \frac{1}{\sqrt{N}}\sum_{\k \in BZ } a_{\k}, \\
    (Va)_{0} &= \frac{1}{\sqrt{N}}\sum_{\k \in BZ} a_{\k} e^{-i\varphi_{\k}}.
\end{align*}
There are analogous expressions for $(Ub)_0$ and $(Vb)_0$. 
Differentiating $\Bar{E}^{0,z}$ with respect to $a_{\k}$ and $b_{\k}$ gives: 
\begin{equation}\label{eq:Minimise_wrt_ak}
\begin{split}
    4\sin^2(\theta/2)&(|\Delta_{\k}|+\Lambda_{a}) a_{\k}^{*} + \Lambda_{ab}b_{\k}^{*} \\
    &= \frac{4}{\sqrt{N}}\sin^2(\theta/2)((Ua)_{0}e^{-i\varphi_{k}} +(Va)_{0}) \\
    &\quad - \frac{2}{\sqrt{N}}\sin(\theta)((Ub)_0e^{-i\varphi_{\k}}-(Vb)_0),
\end{split}
\end{equation}
and
\begin{equation}
\begin{split}
    4\sin^2(\theta/2)&(|\Delta_{\k}|+\Lambda_{b}) b_{\k}^{*} + \Lambda_{ab}a_{\k}^{*}  \\
    &= \frac{4}{\sqrt{N}}\sin^2(\theta/2)((Ub)_0e^{-i\varphi_{k}} +(Vb)_0) \\
    &\quad + \frac{2}{\sqrt{N}}\sin(\theta)((Ua)_{0}e^{-i\varphi_{\k}}-(Va)_{0}).
\end{split}
\end{equation}
Multiplying Eq.~(\ref{eq:Minimise_wrt_ak}) by $b_{\k}$ and summing over the Brillouin zone gives an equation for the Lagrange multiplier $\Lambda_{ab}$: 
\begin{equation}
\begin{split}
    4&\sin^2(\theta/2)\sum_{\k} b_{\k} |\Delta_{\k}| a_{\k}^{*} + \Lambda_{ab} \\
    &= 4\sin^2(\theta/2)((Ua)_{0}(Vb)_0 +(Va)_{0}(Ub)_0).
    \end{split}
\end{equation}
We will seek a solution to the equations with $\Lambda_{ab}=0$. This is possible by exploiting the $U(1)$ redundancy of Eq.~(\ref{eq:U(1)_redundancy}) to require that $(Ua)_{0} = (Va)_{0}$ and $(Ub)_0 = -(Vb)_0$. 
Enforcing this ensures $\sum_{\k} |\Delta_{\k}| b_{\k}^* a_{\k} = 0$ is also satisfied. A solution of this form has: 
\begin{equation}\label{eq:a_k}
    a_{\k} = \frac{A_a}{\sqrt{N}}\frac{1+e^{i\varphi_{\k}}}{\Lambda_{a}+|\Delta_{\k}|},
\end{equation}
and
\begin{equation}\label{eq:b_k}
    b_{\k} = \frac{A_b}{\sqrt{N}}\frac{e^{i\varphi_{\k}}-1}{\Lambda_{b}+|\Delta_{\k}|},
\end{equation}
with
\begin{align}\label{eq:A_a}
    A_a &= (Ua)_{0} - (Ub)_0\cot(\theta/2), \\
    A_b &= (Ub)_0 +(Ua)_{0}\cot(\theta/2). \label{eq:A_b}
\end{align}
 The quantities $A_a$ and $A_b$ may be related to $\Lambda_{a}$ and $\Lambda_{b}$ by using the normalisation constraints. Furthermore, Eq.~(\ref{eq:A_a}) and Eq.~(\ref{eq:A_b}) allow $\Lambda_{b}$ and $\theta$ to be determined from $\Lambda_{a}$. Minimising $\Bar{E}^{0,z}$ with respect to $\theta$ gives 
\begin{equation}\label{eq:Min_wrt_theta}
\begin{split}
    \sin(\theta)\sum_{\k}&|\Delta_{\k}|(|b_{\k}|^2+|a_{\k}|^2) + 4\cos(\theta)(Ua)_{0}(Ub)_0 \\
    &= 2\sin(\theta)((Ua)_{0}^2 - (Ub)_0^2).
\end{split}
\end{equation}
Using Eq.~(\ref{eq:a_k}) and Eq.~(\ref{eq:b_k}), 
\begin{align*}
    \sum_{\k}|\Delta_{\k}||a_{\k}|^2 &= 2A_a(Ua)_{0} - \Lambda_{a},\\
    &= 2(Ua)_{0}^2 -2(Ua)_{0}(Ub)_0\cot(\theta/2) - \Lambda_{a} \\
    \sum_{\k}|\Delta_{\k}||b_{\k}|^2 &= -2A_b(Ub)_0 - \Lambda_{b} \\
    &= -2(Ub)_0^2 -2(Ua)_{0}(Ub)_0\cot(\theta/2) - \Lambda_{b}. 
\end{align*}
Substituting these into Eq.~(\ref{eq:Min_wrt_theta}) and simplifying gives
\begin{align}
    \Lambda_{a} +\Lambda_{b} = -4(Ua)_{0}(Ub)_0\csc(\theta).
\end{align}
Both sides of the equation may be regarded as functions of $\Lambda_{a}$. Solving numerically for $\Lambda_{a}$, and using Eq.~(\ref{eq:A_a}) and Eq.~(\ref{eq:A_b}) to determine $\Lambda_{b}$ and $\theta$ gives: 
\begin{align*}
    \Lambda_{a} = 1.5396, \\
    \Lambda_{b} = 0.2995, \\
    \theta = 0.9561.
\end{align*}
This may be generalised to an arbitrary vison pair site $\r$ and orientation $\alpha$ to give the result stated in Eq.~(\ref{eq:a_k_optimised}) and Eq.~(\ref{eq:b_k_optimised}) in the main text. 

\section{Calculation of hopping parameters}\label{App:Hopping_parameters}

The matrix elements that must be calculated to calculate the hopping parameters of vison pairs have the general form: 
\begin{widetext}
\begin{equation}\label{eq:GS_overlaps}
\braket{0^{\R,\alpha}|0^{\R',\beta}} = \cos^2(\theta/2) + \sin^2(\theta/2)((a^{\R,\alpha}\cdot a^{\R',\beta})(b^{\R,\alpha}\cdot b^{\R',\beta}) - (a^{\R,\alpha}\cdot b^{\R',\beta})(b^{\R,\alpha}\cdot a^{\R',\beta}))
\end{equation}
\begin{equation}\label{eq:2_fermion_matrix_element}
    \begin{split}
    \frac{K}{|K|}\bra{0^{\R,\alpha}}ic_{\r}^{A}c_{\r'}^{B}\ket{0^{\R',\beta}} 
    = &-(U^{u_0}V^{u_{0}T})_{\r,\r'}\braket{0^{\R,\alpha}|0^{\R',\beta}}\\ 
    + \frac{1}{2}\sin(\theta)\Big(&(Ub^{\R,\alpha})_{\r}(Va^{\R,\alpha})_{\r'}-(Vb^{\R,\alpha})_{\r'}(Ua^{\R,\alpha})_{\r}\Big) \\
    + \frac{1}{2}\sin(\theta)\Big(&Ub^{\R',\beta})_{\r}(Va^{\R',\beta})_{\r'}-(Vb^{\R',\beta})_{\r'}(Ua^{\R',\beta})_{\r}\Big) \\
    +\sin^2(\theta/2)\Big(&(b^{\R,\alpha}\cdot b^{\R',\beta})((Ua^{\R,\alpha})_{\r}(Va^{\R',\beta})_{\r'}+(Va^{\R,\alpha})_{\r'}(Ua^{\R',\beta})_{\r}) \\
    +&(a^{\R,\alpha}\cdot a^{\R',\beta})((Ub^{\R,\alpha})_{\r}(Vb^{\R',\beta})_{\r'}+(Vb^{\R,\alpha})_{\r'}(Ub^{\R',\beta})_{\r}) \\
    -&(a^{\R,\alpha}\cdot b^{\R',\beta})((Ub^{\R,\alpha})_{\r}(Va^{\R',\beta})_{\r'}+(Vb^{\R,\alpha})_{\r'}(Ua^{\R',\beta})_{\r})\\
    -&(b^{\R,\alpha}\cdot a^{\R',\beta})((Ua^{\R,\alpha})_{\r}(Vb^{\R',\beta})_{\r'}+(Va^{\R,\alpha})_{\r'}(Ub^{\R',\beta})_{\r})\Big)
    \end{split}
\end{equation}
\text{Where:} 
\begin{align}
    (Ua^{(\R,\alpha)})_{\r} &= \frac{A_{a}}{N}\sum_{\k\in BZ}e^{i\k\cdot(\r-\R)}\frac{1+e^{i(\k\cdot(\r_{z}-\r_{\alpha})-\varphi_{\k})}}{\Lambda_{a}+|\Delta_{\k}|}\\
    (Va^{(\R,\alpha)})_{\r} &= (Ua^{(\R,\alpha)})_{2\R-\r+\r_{\alpha}-\r_{z}}\\
    (Ub^{(\R,\alpha)})_{\r} &= \frac{A_{b}}{N}\sum_{\k\in BZ}e^{i\k\cdot(\r-\R)}\frac{1-e^{i(\k\cdot(\r_{z}-\r_{\alpha})-\varphi_{\k})}}{\Lambda_b+|\Delta_{\k}|} \\
    (Vb^{(\R,\alpha)})_{\r} &= -(Ub^{(\R,\alpha)})_{2\R-\r+\r_{\alpha}-\r_{z}}
\end{align}
\begin{align}
    (a^{(\R,\alpha)}\cdot a^{(\R',\beta)}) &= \frac{A_{a}^{2}}{N}\sum_{\k\in BZ} e^{i\k\cdot(\R-\R')}\frac{1+e^{i\k\cdot(\r_{\alpha}-\r_{z})+\varphi_{\k}}+e^{-i\k\cdot(\r_{\beta}-\r_{z})-\varphi_{\k}}+e^{i\k\cdot(\r_{\alpha}-\r_{\beta})}}{(\Lambda_a+|\Delta_{\k}|)^{2}} \\
    (b^{(\R,\alpha)}\cdot b^{(\R',\beta)}) &= \frac{A_{b}^{2}}{N}\sum_{\k \in BZ} e^{i\k\cdot(\R-\R')}\frac{1-e^{i\k\cdot(\r_{\alpha}-\r_{z})+\varphi_{\k}}-e^{-i\k\cdot(\r_{\beta}-\r_{z})-\varphi_{\k}}+e^{i\k\cdot(\r_{\alpha}-\r_{\beta})}}{(\Lambda_{b}+E_{\k})^2} \\
    (a^{(\R,\alpha)}\cdot b^{(\R',\beta)}) &= \frac{A_{b}A_{a}}{N}\sum_{\k \in BZ} e^{i\k\cdot(\R-\R')}\frac{1+e^{i\k\cdot(\r_{\alpha}-\r_{z})+\varphi_{\k}}-e^{-i\k\cdot(\r_{\beta}-\r_{z})-\varphi_{\k}}-e^{i\k\cdot(\r_{\alpha}-\r_{\beta})}}{(\Lambda_{a}+|\Delta_{\k}|)(\Lambda_{b}+|\Delta_{\k}|)} 
\end{align}
\end{widetext}
\subsection{Heisenberg Interaction}
The Heisenberg interaction Hamiltonian is given by: 
\begin{align*}
    \mathcal{H}^{J} &= J\sum_{\substack{\r \in A\\
    \alpha = z,y,z}} \sigma_{\r}\cdot \sigma_{\r+\r_{\alpha}}
\end{align*}
Using the expression from Ref.~\cite{zhang_variational_2021}, the hopping parameter for vison pairs in the presence of Heisenberg interactions:
\begin{align*}
    \mathcal{T}^{J} &= \bra{\R+\a_1,z}\mathcal{H}^{J}\ket{\R,z} \nonumber\\
    &= J\big(\braket{{0^{\R+\a_1,z}}|{0^{\R,z}}}+\bra{0^{\R+\a_1,z}}ic_{\R+\a_1}^{A}c_{\R}^{B}\ket{0^{\R,z}}\big)
\end{align*}
Using the general formulae Eq.~(\ref{eq:GS_overlaps}) and Eq.~(\ref{eq:2_fermion_matrix_element}): 
\begin{align*}
    \braket{{0^{\R+\a_{1},z}}|{0^{\R,z}}}&= 0.7977 \\
    \bra{0^{\R+\a_{1},z}}ic_{\R+\a_{1}}^{A}c_{\R}^{B}\ket{0^{\R,z}} &= -0.6986\sign(K)
\end{align*}

\subsection{Gamma Interactions}
The Gamma interaction Hamiltonian is given by:
\begin{equation}
    \mathcal{H}^{\Gamma} = \Gamma\sum_{\substack{\r \in A \\\alpha\neq\beta\neq\gamma}} \sigma_{\r}^{\beta}\sigma_{\r+\r_{\alpha}}^{\gamma}
\end{equation}
The hopping parameter for vison pairs due to Gamma interactions~\cite{zhang_variational_2021} is:
\begin{align}
    \mathcal{T}^{\Gamma} &= \Gamma\big(\bra{0^{\R+\a_1,y}}ic_{\R+\a_{1}}^Ac_{\R}^B\ket{0^{\R,z}}-\braket{0^{\R+\a_{1},y}|0^{\R,z}}\big)
\end{align}
Using the general formulae Eq.~(\ref{eq:GS_overlaps}) and Eq.~(\ref{eq:2_fermion_matrix_element}): 
\begin{align*}
    \braket{{0^{\R+\a_{1},y}}|{0^{\R,z}}}&= 0.8220 \\
    \bra{0^{\R+\a_{1},y}}ic_{\R+\a_{1}}^{A}c_{\R}^{B}\ket{0^{\R,z}} &= -0.6411\sign(K)
\end{align*}

\subsection{Zeeman Interactions}
The Zeeman interaction for a uniform magnetic field $h$ in the $[1,1,1]$ direction is given by: 
\begin{align}
    \mathcal{H}^{h} = h\sum_{\substack{\r \in A,B\\\alpha = x,y,z}}\sigma_{\r}^{\alpha}
\end{align}
The hopping parameter for vison pairs due to the Zeeman interaction at first order is given by:
\begin{equation}
    \mathcal{T}^{h} = ih\big(\braket{0^{\R,\beta}|0^{\R,\alpha}}-\bra{0^{\R,\beta}}ic_{\R}^{A}c_{\R+\r_{\gamma}-\r_{z}}^{B}\ket{0^{\R,\alpha}}\big)
\end{equation}
Note that in this calculation we have neglected the change to the matter sector Hamiltonian due to the Haldane mass term which is induced at third order in the magnetic field strength \cite{kitaev_anyons_2006}. Inclusion of these effects is open for further work. 
Using the general formulae Eq.~(\ref{eq:GS_overlaps}) and Eq.~(\ref{eq:2_fermion_matrix_element}): 
\begin{align*}
    \braket{0^{\R,\beta}|0^{\R,\alpha}}&= 0.7184 \\
    \bra{0^{\R,\beta}}ic_{\R}^{A}c_{\R+\r_{\gamma}-\r_{z}}^{B}\ket{0^{\R,\alpha}} &= -0.6689\sign(K)
\end{align*}

\bibliography{constructing_variational_ground_states.bib}

\end{document}